# ROLE OF AZIMUTHAL ENERGY FLOWS IN THE GEOMETRIC SPIN HALL EFFECT OF LIGHT

A.Ya. Bekshaev

*I.I. Mechnikov National University, Dvorianska 2, 65082, Odessa, Ukraine*
*E-mail address*: bekshaev@onu.edu.ua

**Abstract**

In an oblique section of a paraxial beam with angular momentum, the beam center of gravity (CG) is shifted with respect to its position in the normal cross section. We relate this shift with the internal energy redistribution occurring on the passage between the normal and oblique sections. The transverse orbital flow explains the effect for scalar beams, similar incorporation of the spin flow enables explanation of the CG shift in oblique sections of the elliptically polarized beams. Role of the special properties of the position-sensitive detector placed in the oblique beam section is discussed.

During the past decades, a substantial attention is being paid to the interrelations between the light beam trajectory and its intrinsic structure [1–14]. Usually, the beam trajectory is expressed by the "center of gravity" (CG) for the transverse distribution of the beam energy density $w$ or the longitudinal component of the energy flow density $S_z$ that in paraxial approximation practically coincide [2]:

$$\mathbf{r}_c(z) = \begin{pmatrix} x_c(z) \\ y_c(z) \end{pmatrix} = \frac{\int \mathbf{r} w(\mathbf{r},z) d^2\mathbf{r}}{\int w(\mathbf{r},z) d^2\mathbf{r}} \approx \frac{\int \mathbf{r} S_z(\mathbf{r},z) d^2\mathbf{r}}{\int S_z(\mathbf{r},z) d^2\mathbf{r}} \quad (1)$$

($\mathbf{r}$ is the transverse radius-vector, the beam is supposed to propagate along axis $z$ and integration over the whole beam cross section is implied). The intrinsic structure of a monochromatic paraxial beam with frequency $\omega$ and wave number $k$ is characterized by the slowly varying complex amplitude $\mathbf{u}(\mathbf{r},z)$ which determines the electric and magnetic fields of the beam [2]

$$\mathbf{E}(\mathbf{r},z) = \mathbf{E}_\perp + \mathbf{e}_z E_z = \left[\mathbf{u} + \frac{i}{k}\mathbf{e}_z(\nabla_\perp \cdot \mathbf{u})\right]e^{ikz}, \quad (2)$$

$$\mathbf{H}(\mathbf{r},z) = \mathbf{H}_\perp + \mathbf{e}_z H_z = \left[(\mathbf{e}_z \times \mathbf{u}) + \frac{i}{k}\mathbf{e}_z(\nabla_\perp \cdot (\mathbf{e}_z \times \mathbf{u}))\right]e^{ikz}, \quad (3)$$

so the "true" instantaneous electric field in the beam can be found as $\mathrm{Re}\left[\mathbf{E}(\mathbf{r},z)\exp(-i\omega t)\right]$. In the paraxial limit [2]

$$w(\mathbf{r},z) = \frac{1}{c}S_z(\mathbf{r},z) = \frac{1}{8\pi}(\mathbf{u}^* \cdot \mathbf{u}) \quad (4)$$

where $c$ is the light velocity and the Gaussian system of units is employed.

A classical example of the structure-induced trajectory modification is supplied by the known Imbert-Fedorov shift [3–5, 8–12] that occurs when an elliptically polarized beam falls onto a boundary between two homogeneous dielectric media: the CGs of the reflected and refracted beams experience shifts orthogonal to the plane of incidence and depending on the polarization ellipticity (spin Hall effect of light). Further, similar effects associated with the transverse inhomogeneity of scalar beams have been predicted and detected [10, 13–17]. Thorough theoretical investigations have disclosed common nature of the mentioned effects and their deep relation to the vortex properties of the beam expressed by its spin and/or orbital angular momentum (AM) [1, 2, 10, 18–21]. Actually, the observable shift of the beam CG consists of two constituents: the "material" contribution depends on optical properties of the contacting media and of their interface whereas the "geometric" one is determined completely by mutual disposition of the boundary and the beam [16, 17, 22].

In its pure form, the geometric effect manifests itself in an oblique section of a paraxial beam [22, 23]. Its physical nature can be illustrated by Figs. 1 and 2. The presence of the AM in the incident beam is usually accompanied by certain azimuthal component of the energy flow which is shown by the bent arrow in Fig. 1. Initially, the beam energy is uniformly distributed along each transverse circumference centred at the beam axis $z$. This is schematically depicted by equidistant ray positions in the initial cross section (Fig. 2, open circles). Then, each portion of the light energy propagating along the $z$ axis, simultaneously experiences certain azimuthal displacement, and this continues until the corresponding ray meets the oblique plane $(x', y)$. But the "upper side" (in Fig. 1; in Fig. 2 its counterpart lies to the left of the beam axis) traverses longer distance than its "lower side" ($\Delta z$ in Fig. 1). As a result, rays of the upper side experience larger azimuthal displacements; this causes that the rays accumulate in the lateral part of the $(x', y)$ plane (lower part in Fig. 2). The beam energy effectively shifts orthogonally to the plane of incidence as is shown by the black arrow in Fig. 1.

Quantitative characteristics of this shift depend on the incident beam complex amplitude distribution and on the properties of a photodetector placed in the $(x', y)$ plane: whether it is sensitive to the total field energy, or to a certain polarization component or to a certain component of the energy flow density **S**, whether it completely absorbs the incident light energy or partially reflects it, etc. In case of a totally absorbing detector measuring the light energy density and for the scalar beam model, the lateral CG shift was shown to be equal [22]

$$\Delta y_c = m_{xy} \tan\theta \tag{5}$$

where $\theta$ is the angle of incidence, $m_{xy}$ is the element of the moment matrix [24] that, in accord to [22], characterizes the transverse energy redistribution in the beam:

$$\mathsf{M}_{12} = \begin{pmatrix} m_{xx} & m_{xy} \\ m_{yx} & m_{yy} \end{pmatrix} = \frac{1}{\Phi} \int \begin{pmatrix} xS_x & xS_y \\ yS_x & yS_y \end{pmatrix} d^2\mathbf{r} \ . \tag{6}$$

Here $S_x$ and $S_y$ are the transverse components of the beam Poynting vector (energy flow density),

$$\Phi = \int S_z(\mathbf{r}) d^2\mathbf{r} = \frac{c}{8\pi} \int |u|^2 d^2\mathbf{r} \tag{7}$$

is the total beam power (longitudinal energy flow). Note that relations (5) and (6) were derived for scalar beams, so that $S_x$ and $S_y$ in (6) represent the orbital flow density [25, 26] owing to the spatial inhomogeneoty of the scalar complex amplitude.

Another and more general version of the geometric spin Hall effect was described by A. Aiello et al. [23]. The authors define the CG position via distribution of the energy flow density component normal to the oblique section,

$$\mathbf{r}'_c = \begin{pmatrix} x'_c \\ y'_c \end{pmatrix} = \frac{\int \mathbf{r}' S_{z'}(\mathbf{r}') d^2\mathbf{r}'}{\int S_{z'}(\mathbf{r}') d^2\mathbf{r}'} \ , \tag{8}$$

and show that, if the incident beam possesses a non-zero AM, the CG (8), measured in the oblique section, appears to be shifted orthogonally to the incidence plane. This shift is proportional to the $x'$-component of the incident beam AM regardless of its nature (spin, orbital, intrinsic or extrinsic). In particular, for a homogeneously polarized Gaussian beam with

$$\mathbf{u} = \hat{\mathbf{x}} u_X + \hat{\mathbf{y}} u_Y = (\hat{\mathbf{x}}\alpha + \hat{\mathbf{y}}\beta) u \ , \tag{9}$$

where $\hat{\mathbf{x}}$, $\hat{\mathbf{y}}$ are the unit vectors of the coordinate axes,

$$u = \frac{1}{1 + iz/z_R} \exp\left[-\frac{x^2 + y^2}{2b_0^2(1 + iz/z_R)}\right],$$

the CG shift calculated in [23] is

$$\Delta y'_c = \frac{\sigma}{2k}\tan\theta \qquad (10)$$

where

$$\sigma = i\left(\alpha\beta^* - \alpha^*\beta\right) \qquad (11)$$

is the polarization ellipticity.

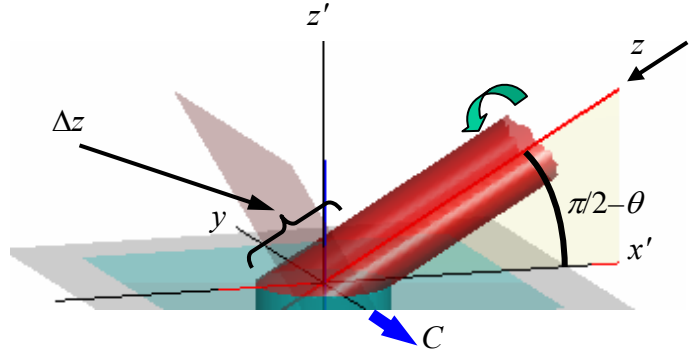

Fig. 1. Transverse shift of the beam impinging an oblique detector, which is caused by the circular energy flow (see explanations in text).

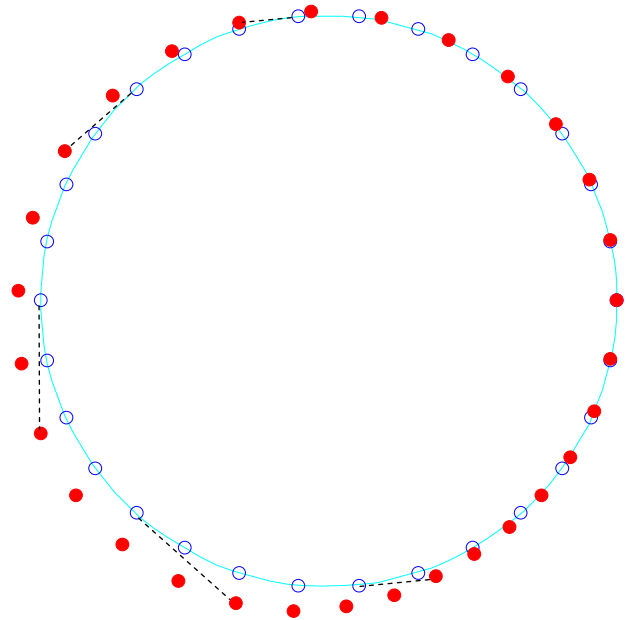

Fig. 2. Schematic redistribution of rays in the beam due to interaction with the oblique plane (projections on the incident beam cross section when seeing along its axis). Open circles: initial positions; filled circles: points at which rays meet the oblique plane $(x', y)$ of Fig. 1. Some azimuthal displacements are shown by dashed lines.

The conclusion of [23] is quite general but rather formal and leaves unclear the physical reasons and origination of the effect, which gives the authors of [23] a right to qualify it as a "counter-intuitive" phenomenon. In the rest of this note, we intend to demonstrate that, on the contrary, the lateral displacement of the beam CG in the oblique section quite agrees with intuitive arguments based on the vivid picture of the transverse energy circulation in the incident beam. Simultaneously, we will show that in practical situations it is not the incident beam AM but rather the internal energy flow that plays the decisive role in this effect.

Let us consider a paraxial incident beam satisfying Eq. (9). In what follows, we will not require the beam to be Gaussian but suppose it to be homogeneously polarized and axially symmetric so that the transverse orbital flow vanishes. However, the spin flow can exist whose density is described by formula (9) of [25]:

$$\mathbf{S}_C = \frac{1}{2k}\left(\hat{\mathbf{x}}\frac{\partial s_3}{\partial y} - \hat{\mathbf{y}}\frac{\partial s_3}{\partial x}\right) \quad (12)$$

where

$$s_3(\mathbf{r},z) = \frac{ic}{8\pi}\left(u_X u_Y^* - u_X^* u_Y\right) = \frac{c}{8\pi}\sigma|u|^2 \quad (13)$$

is the Stokes parameter responsible for the degree of circular polarization. The second assumption is that the spin flow produces the beam CG shift in the oblique section in the manner analogous to the orbital flow action (Eqs. (5), (6)):

$$\Delta y_c^C = m_{xy}^C \tan\theta, \quad m_{xy}^C = \frac{1}{\Phi}\int x S_{Cy}\, d^2\mathbf{r}. \quad (14)$$

Then, from Eqs. (9) and (11) – (13), using the integration by parts with taking into account that function $s_3(x,y)$ vanishes at the transverse infinity, we easily find

$$m_{xy}^C = \frac{1}{2k\Phi}\int x\left(-\frac{\partial s_3}{\partial x}\right)d^2\mathbf{r} = \frac{1}{2k\Phi}\int s_3(\mathbf{r})d^2\mathbf{r} = \frac{1}{2k\Phi}\frac{c}{8\pi}\sigma\int|u|^2 d^2\mathbf{r} = \frac{1}{2k}\sigma \quad (15)$$

after which first Eq. (14) ultimately gives

$$\Delta y_c^C = \frac{1}{2k}\sigma\tan\theta, \quad (16)$$

which exactly coincides with the result (10) obtained via the formal consideration in [23]. However, (16) differs from (10) in two important aspects:
(i) Eqs. (5), (6) and, consequently, (14) and (16) are derived for the electromagnetic energy distribution while Eq. (10) pertains to the energy flow component $S_{z'}$ normal to the oblique section;
(ii) Eq. (16) expresses the result expected from an absorbing detector placed in the oblique plane whereas Eq. (10) represents a rather abstract quantity whose measuring procedure is not obvious.

Anyway, Eqs. (14) and (16) show that the lateral shift of the beam CG when it falls onto an oblique surface (oblique-incidence geometric spin Hall effect) can be explained by the azimuthal energy flow, no matter of the "spin" or "orbital" origination. On the other hand, the beam AM can also be related to the azimuthal energy flow, and that is where the connection between the AM and the lateral beam shift comes from. The direct link between $\Delta y_c$ and the incident beam AM, substantiated in [23], is rather formal and can be realized only in certain special situations. Furthermore, in some practical cases the lateral shift of this type can take place even when the incident beam possesses no AM (e.g. in scalar beams when in matrix (6) $m_{xy} = m_{yx} \neq 0$ [27]) and, otherwise, a beam with non-zero AM may show no lateral shift ($m_{xy} = 0$, $m_{yx} \neq 0$). Maybe, the simplest examples for both cases are provided by the astigmatic Gaussian beams with the complex amplitude distribution

$$u(x,y) = \exp\left[-\frac{1}{2}\left(b_{xx}x^2 + b_{yy}y^2 + 2b_{xy}xy\right)\right]\exp\left[i\frac{k}{2}\left(a_{xx}x^2 + a_{yy}y^2 + 2a_{xy}xy\right)\right]. \quad (17)$$

For simplicity, let the beam be linearly polarized ($\sigma = 0$). Then in agreement to Eqs. (9) and (10) of [27]

$$m_{xy} = \frac{b_{yy}a_{xy} - b_{xy}a_{yy}}{2B}, \quad m_{yx} = \frac{b_{xx}a_{xy} - b_{xy}a_{xx}}{2B}, \quad B = b_{xx}b_{yy} - b_{xy}^2, \quad (18)$$

the spin AM vanishes, the whole beam AM is orbital and, due to Eq. (19) of [22], in units $\hbar$ per photon this AM equals to

$$\Lambda = k\left(m_{xy} - m_{yx}\right) = k\frac{b_{xy}\left(a_{xx} - a_{yy}\right) - a_{xy}\left(b_{xx} - b_{yy}\right)}{2B}. \quad (19)$$

Now it is not difficult to specify both situations more accurately.

1. $\Lambda = 0$, $m_{xy} \neq 0$ (lateral shift without AM) occurs, for example, in beam (17) with $b_{xx} = b_{yy} = b$, $b_{xy} = 0$. Then, in accord with (5) and (19),

$$m_{xy} = m_{yx} = \frac{1}{2b}\alpha_{xy}, \quad \Lambda = 0, \quad \Delta y_c = \frac{\alpha_{xy}}{2b}\tan\theta. \tag{20}$$

2. $\Lambda \neq 0$, $m_{xy} = 0$ (AM without lateral shift) is realized if $m_{xy} = 0$ holds together with $m_{yx} \neq 0$. For example, if for given $b_{xy}$, $b_{yy}$, and $a_{yy}$ we choose $a_{xy} = \frac{b_{xy}a_{yy}}{b_{yy}}$, requirement $m_{xy} = 0$ is satisfied and, consequently, $\Delta y_c = 0$. At the same time, there is enough freedom for choosing other parameters so that

$$\Lambda = -km_{yx} = -k\frac{b_{xy}}{b_{yy}}\frac{b_{xx}a_{yy} - a_{xx}b_{yy}}{2B} \neq 0. \tag{21}$$

Of course, all the above precautions apply only for beams with complicated spatial structure, so the conclusions of [23], directly addressing the beams with elliptic polarization, are still correct.

In conclusion, we would like to emphasize the two points addressed in this note. First, it is the assumption that the spin flow produces the transverse energy redistribution in the manner analogous to the orbital flow action described by Eqs. (5) and (6). In fact, Eqs. (14) were not derived from special properties of vector paraxial fields but were just written by analogy. On the other hand, there are some vague aspects as to the spin flow behavior and even its observability when the beam interacts with obstacles [2] (to which category the oblique detector can be assimilated). Consequently, if the discussed effect will be discovered, e.g., for the circularly polarized Gaussian beam falling onto an oblique detector, it will be evidence that the spin flow really exist and can be observed.

Second, our consideration is essentially based on the nature of the photodetector used for the beam CG localization. This leads to generalization that any manifestation of the geometric spin Hall effect of light essentially relies upon the special ways of the light detection and, furthermore, of the light interaction with the oblique plane in which the effect is expected. Anyway, the effect is not purely "geometric" and includes many physical aspects that can substantially modify its manifestation down to its complete vanishing.